\documentclass[aps,prb,twocolumn,nofootinbib,showpacs,preprintnumbers,notitlepage,groupedaddress]{revtex4-2}

\usepackage{graphicx}
\usepackage{graphics}
\usepackage{amsmath}
\usepackage{amssymb}
\usepackage{amsfonts}
\usepackage{dsfont}
\usepackage{braket}
\usepackage{color}
\usepackage{braket,slashed}
\usepackage[mathscr]{euscript}
\definecolor{darkblue}{rgb}{0, 0, 0.8}
\usepackage[colorlinks=true, breaklinks=true, linkcolor=red, citecolor=blue, urlcolor=blue]{hyperref} 
\usepackage{hyperref}
\usepackage{subfigure}
\usepackage{xfrac}
\usepackage{bm}
\usepackage{kantlipsum}
\usepackage{enumitem}
\usepackage{tikz}
\usepackage{framed}
\usepackage{graphicx}
\usepackage{subfigure}
\usepackage{cleveref}

\allowdisplaybreaks[1]

\newcommand{\code}[1]{\texttt{#1}}

\renewcommand{\d}{\ensuremath{\mathrm{d}}}
\newcommand{\e}{\ensuremath{\mathrm{e}}}

\newcommand{\Z}{\ensuremath{\mathbb{Z}}}

\newcommand{\Heff}{\ensuremath{H_{\mathrm{eff}}}}

\begin{document}

\title{Momentum-resolved time evolution with matrix product states}

\author{Maarten Van Damme}
\email{Maarten.VanDamme@ugent.be}
\author{Laurens Vanderstraeten}
\email{Laurens.Vanderstraeten@ugent.be}

\affiliation{Department of Physics and Astronomy, University of Ghent, Krijgslaan 281, 9000 Gent, Belgium}

\begin{abstract}
We introduce a method based on matrix product states (MPS) for computing spectral functions of (quasi) one-dimensional spin chains, working directly in momentum space in the thermodynamic limit. We simulate the time evolution after applying a momentum operator to an MPS ground state by working with the momentum superposition of a window MPS. We show explicitly for the spin-1 Heisenberg chain that the growth of entanglement is smaller in momentum space, even inside a two-particle continuum, such that we can attain very accurate spectral functions with relatively small bond dimension. We apply our method to compute spectral lineshapes of the gapless XXZ chain and the square-lattice $J_1$-$J_2$ Heisenberg model on a six-leg cylinder.
\end{abstract}

\maketitle

\newcommand{\diagram}[1]{\;\vcenter{\hbox{\includegraphics[scale=0.32,page=#1]{./diagrams_main.pdf}}}\;}

\par\noindent\emph{\textbf{Introduction---}} %
Spectral functions are one of the most important tools for relating theory and experiment in the field of strongly-correlated quantum matter. The most exotic properties of strongly-correlated quantum phases are typically related to the low-energy excitations in these systems, for which the spectral function is a direct experimental probe. Inelastic neutron scattering for magnetic materials and angle-resolved photo-emission spectroscopy for electronic systems serve as the most interesting options. As an outstanding example, in the last few years, the occurrence of fractionalized quasiparticles in candidate two-dimensional spin-liquid materials has been observed through the spectral function in neutron-scattering experiments \cite{Han2012, Banerjee2016}.
\par A pressing challenge on the theoretical side is the evaluation of the spectral function for a given microscopic model Hamiltonian. In the one-dimensional case, only a restricted class of integrable models such as the spin-1/2 Heisenberg chain \cite{Mourigal2013} allows for a quasi-exact computation of the spectral function. In two dimensions, besides the case of the Kitaev spin liquid \cite{Knolle2014, Knolle2015}, exact calculations are even less common. Traditional numerical methods include exact diagonalization with a continued fraction expansion \cite{Gagliano1987}, which is limited to small clusters, and quantum Monte Carlo, for which the analytic continuation \cite{Schuttler1986, Sandvik1998, Shao2017} to real frequencies is often uncontrolled. Variational methods for capturing excited states can be based on Gutzwiller-projected mean-field states \cite{Li2010, Ferrari2018} or, more recently, variational wavefunctions from machine learning \cite{Hendry2019}.
\par For one-dimensional lattice systems, a lot of effort went into designing efficient numerical methods based on the formalism of matrix product states (MPS) \cite{Schollwoeck2011}. The earliest approaches were based on the continued-fraction expansion \cite{Hallberg1995, Dargel2011} or correction vectors \cite{Kuhner1999, Jeckelmann2002, Benthien2004}; the latter can give very accurate results, but require a different run for every frequency and adding a small imaginary frequency leading to significant broadening. Using Chebyshev expansions for the spectral function, the full frequency range can be computed in a single run \cite{Holzner2011, Wolf2015, Xie2018}. Alternatively, one can use time-dependent MPS algorithms \cite{White2004, Daley2004, Haegeman2011, Zaletel2015, Paeckel2019} for computing the real-time evolution after applying a local operator to the ground state, and transform to frequency space \cite{White2008}; here, the frequency resolution is limited due to the growth of entanglement in the time-evolved state, although linear prediction \cite{White2008, Pereira2008, Barthel2009} or recursion methods \cite{Tian2021} can be used to extrapolate the signal to longer times.
\par Although the spectral function is a momentum-resolved quantity, in all these MPS approaches the translation symmetry of the model is explicitly broken either by working on a finite open-boundary system or by introducing a local operator for time evolution. A viable alternative consists of targeting the low-energy excitations by a variational ``quasiparticle ansatz'' that has well-defined momentum directly in the thermodynamic limit \cite{Haegeman2012, Haegeman2013}. With this approach, one can compute the dispersion and spectral weight of isolated branches in the spectrum \cite{Haegeman2013b} with high precision \cite{Bera2017, VanDamme2021}; this method does not suffer from a finite resolution in momentum and frequency but only from variational errors. Extending this framework to two-particle excitations \cite{Vanderstraeten2014, Vanderstraeten2015a}, the fine-grained features of the spectral function within a two-particle continuum can be resolved \cite{Vanderstraeten2016}. For critical systems, however, the absence of isolated lines or the occurrence of multi-particle continua make this variational approach less useful for obtaining the full spectral function accurately.
\par In this paper, we introduce an MPS-based method for simulating time evolution directly in momentum space and the thermodynamic limit, which enables us to compute spectral functions with higher accuracy than a real-space approach would with the same bond dimension. We first recapitulate how the real-space approach works using a window MPS embedded in a uniform background, and then discuss how to lift this approach to momentum space. We illustrate our method by comparing the real- and momentum-space approaches for the spectral function in the two-magnon continuum in the spin-1 Heisenberg chain. We further illustrate the efficiency of our momentum-resolved method by simulating the time evolution of wavepackets, thus interpolating between real and momentum space. Finally, we benchmark our method by computing spectral lineshapes for the gapless XXZ chain and the square-lattice $J_1$-$J_2$ Heisenberg model on a six-leg cylinder.

\par\noindent\emph{\textbf{Real-space approach---}} %
For concreteness, let us consider the spin-1 Heisenberg chain with Hamiltonian
\begin{equation}
H = \sum_{\braket{ij}} \vec{S}_i \cdot \vec{S}_j, \qquad \vec{S}_i = \left( S_i^x, S_i^y,S_i^z \right),
\end{equation}
where $S_i^\alpha$ are the spin-1 operators at site $i$ and the sum runs over all nearest neighbors. Here we are interested in the momentum-frequency resolved spectral function at zero temperature
\begin{equation}
S(q,\omega) = \int_{-\infty}^{\infty} \d t \; \e^{i\omega t} S(q,t),
\end{equation}
with
\begin{multline}
S(q,t) = \sum_{\alpha, n'} \e^{-iqn'}  \bra{\Psi_0} S^\alpha_{n+n'} \e^{-i(H-E_0)t} S^\alpha_n \ket{\Psi_0},
\end{multline}
and $\ket{\Psi_0}$ the ground state with ground-state energy $E_0$.
\par First, we approximate the ground state directly in the thermodynamic limit with a uniform MPS described by a single tensor $A$
\begin{equation}
\ket{\Psi_0} = \cdots \diagram{1} \cdots.
\end{equation}
Starting from the ground state, we apply a single-site spin operator at an arbitrary site $n$ and time-evolve the state,
\begin{equation}
\ket{\Psi^\alpha_n(t)}_{\mathrm{rs}} =  \e^{-i(H-E_0)t} S^\alpha_n \ket{\Psi_0}.
\end{equation}
The perturbation induced by the operator is expected to spread through the system as time evolves, such that this time-evolved state should be approximated as an MPS with a window of site-dependent tensors around site $n$, embedded within the uniform MPS \cite{Phien2012, Milsted2013, Zauner2015}
\begin{equation}
\ket{\Psi_n(X)}_{\mathrm{rs}} =  \dots \diagram{2} \cdots.
\end{equation}
In this ansatz, the bond dimension of the tensors $X_i$ and the size of the window $N_w$ enter as control parameters, and both are expected to grow as time evolves: the effect of applying a local operator will spread through the system with a characteristic velocity and the entanglement due to the spread of quasiparticles will grow.

\begin{figure}
\includegraphics[width=\columnwidth]{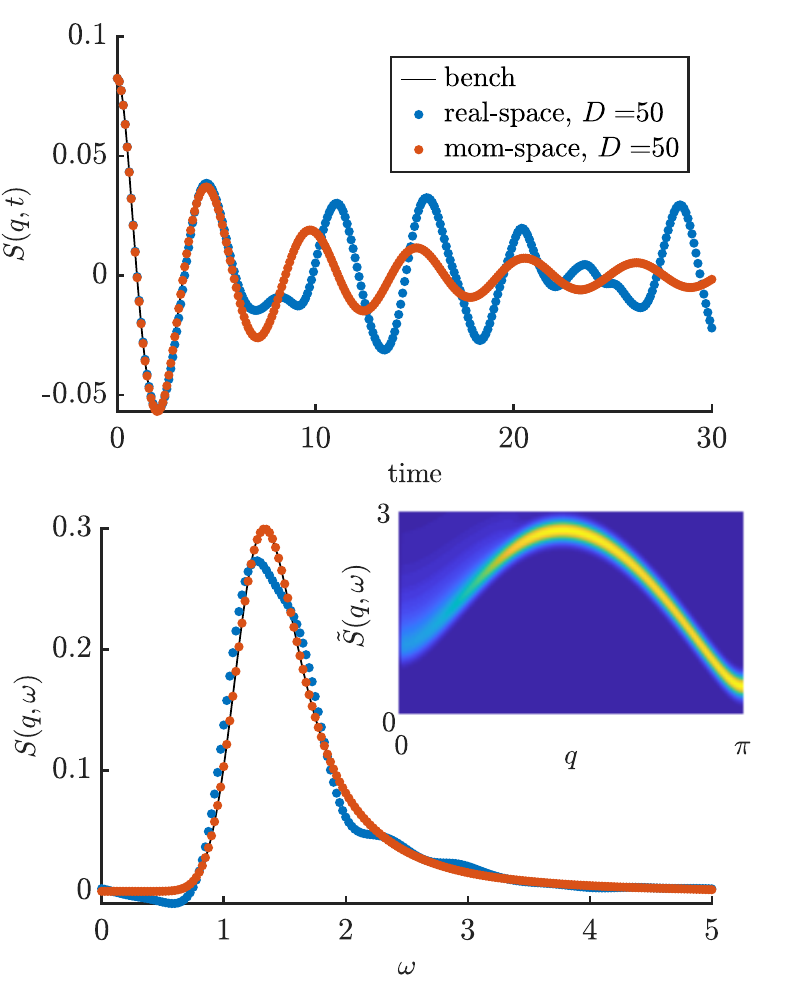}
\caption{\textbf{Comparing the accuracy of real-space and momentum-space approaches.} We plot the real-time signal $S(q,t)$ and the resulting spectral function $S(q,\omega)$ for the momentum cut $q=\pi/10$ inside the two-magnon continuum. We have used SU(2)-symmetric MPS with total bond dimension $D=50$, and one-sided time evolution for the sake of comparison. The benchmark result was run  with a real-space window MPS at a sufficiently large bond dimension $D\approx1500$. The inset shows the rescaled spectral function $\tilde{S}(q,\omega)=S(q,\omega)/\int\d\omega S(q,\omega)$ in the full Brillouin zone, showing the magnon mode and the two-magnon continuum.}
\label{fig:spin1_bond}
\end{figure}

\par The time evolution of the tensors $X_i$ can be determined via the time-dependent variational principle (TDVP) \cite{Haegeman2011, Haegeman2016}, by a Trotter-Suzuki decomposition \cite{Vidal2004, White2004}, or by applying the time-evolution operator in the form of a matrix-product operator (MPO) \cite{Zaletel2015, Vanhecke2021a} and truncating the bond dimension \cite{Paeckel2019}. Here, we take the latter option, i.e. in each time step we want to find new tensors $X_i'$ such that
\begin{equation}
\ket{\Psi_n(X')}_{\mathrm{rs}} \approx \e^{-i(H-E_0)\delta t} \ket{\Psi_n(X)}_{\mathrm{rs}}
\end{equation}
for a time step $\delta t$. This equation can be interpreted variationally such that we find tensors $X_i'$ that maximize the overlap with the right-hand side, which in practice is achieved by a sweeping algorithm that sequentially optimizes over the different tensors; here, exploiting the canonical form and changing the orthogonality center within the window is crucial for a stable algorithm \cite{Schollwoeck2011}. Finally, the overlap of the time-evolved state with a local operator at different sites can be taken and transforming to momentum space yields an estimate of $S(q,t)$.
\par Note that we can split up the full time evolution into two parts such that bra and ket states can be time-evolved independently; this allows us to reach longer times for $S(q,t)$ accurately with a given bond dimension in both time-evolved states. In this work, we always simulate the time evolution up to a total time $T$ and refrain from using extrapolation techniques to focus on the accuracy of the time-evolution method itself. To avoid unphysical signatures of the finite-time simulation in the spectral function, we always apply a Gaussian envelope of the form $\exp(-\alpha t^2/T^2)$ to the finite-time signal before transforming to frequency space.

\par\noindent\emph{\textbf{Momentum-space evolution---}} %
Let us now explain how to evaluate the spectral function directly in momentum space. Here, we start from rewriting the spectral function as an overlap between two states with well-defined momentum
\begin{equation}
S(q,t) =  \sum_\alpha \braket{\Psi_q^\alpha(0) | \Psi_q^\alpha(t)}_{\mathrm{finite}},
\end{equation}
with
\begin{equation}
\ket{\Psi^\alpha_q(t)} =  \e^{-i(H-E_0)t} \sum_n \e^{iqn} S^\alpha_n \ket{\Psi_0},
\end{equation}
and where we have defined the ``finite part'' of the overlap between two momentum states as
\begin{equation}
\braket{\Psi_p | \Psi_q} = 2\pi\delta(p-q) \braket{\Psi_p | \Psi_q}_{\mathrm{finite}}.
\end{equation}
The central idea in this work is to represent this time-evolved momentum state as a ``momentum-window MPS'' \cite{Vanderstraeten2020} of the form
\begin{multline}
\ket{\Psi_q(X)}_{\mathrm{ms}} = \\ \sum_n \e^{iqn} \left( \cdots \diagram{3} \cdots \right).
\end{multline}
Like for the real-space window, the size of the window $N_w$ and the bond dimension of the tensors $X_i$ enter as control parameters. The real-time correlator $S(q,t)$ is now obtained as the overlap of the time-evolved state with the initial one; again, we can perform a two-sided time evolution to reach longer times with the same bond dimension.
\par For a momentum cut that is dominated by a single mode, we know that this momentum-window MPS will capture the time evolution very well. Indeed, the variational quasiparticle ansatz is contained within the class of momentum-window states, so in all cases for which the former yields high-precision eigenstates (i.e., isolated lines in the spectrum \cite{Haegeman2013b}) we will be able to simulate long time evolution accurately with a momentum-window state. For two-particle states, we have captured the ``center-of-mass'' motion by going to momentum space, whereas the relative motion between the two particles must be accounted for by growing the size of the momentum window. In addition to smaller windows, we expect that the bond dimension within the window will grow less rapidly than in the real-space approach. Indeed, since the translation symmetry of the model is explicitly preserved throughout the calculations, we can target one specific momentum sector only. This means that the time-evolved momentum window contains a lot less information, and can therefore be represented with a smaller bond dimension. 
\par Again, the momentum-window MPS can be evolved through time using the TDVP algorithm or applying the evolution operator variationally. In the Appendix we show how to implement these algorithms and we find that the latter option is more efficient; the overhead of working with a momentum superposition is linear in the window size and can be parallelized, so the computational complexity of momentum-resolved time evolution is modest with respect to the real-space version.
\par We can illustrate the efficiency of the momentum-window state by applying it to the spin-1 chain. We find that we can simulate the real-time correlator $S(q,t)$ at momentum $q=\pi$ with high precision up to arbitrary large times with moderate bond dimension (not shown), which follows directly from the fact that the quasiparticle ansatz yields quasi-exact results for the magnon mode \cite{Haegeman2011}. The two-magnon continuum is potentially more challenging: In Fig.~\ref{fig:spin1_bond} we show the results for a momentum cut in the two-particle region $q=\pi/10$, where we compare $S(q,t)$ as evaluated with a real-space and momentum-space window. We have taken sufficiently large window sizes and the same bond dimension, showing clearly that the real-space approach becomes inaccurate rather quickly whereas the momentum-space approach is still accurate at long times. In the spectral function $S(q,\omega)$ the difference is equally clear.

\begin{figure}
\includegraphics[width=\columnwidth]{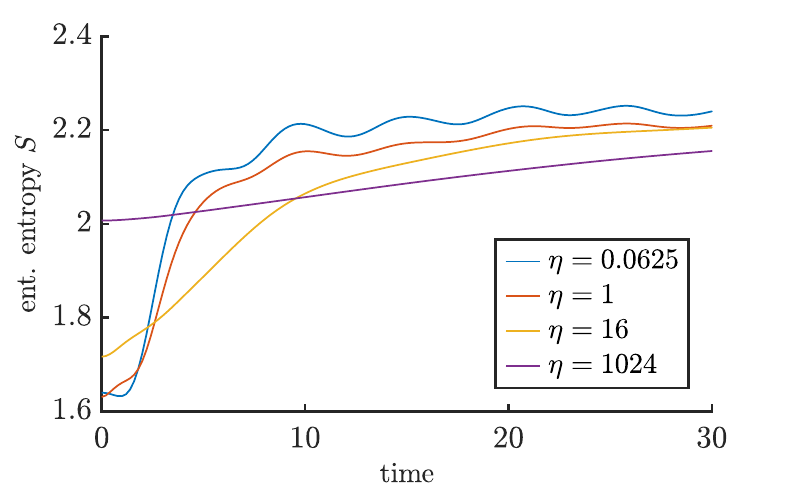}
\caption{\textbf{Interpolating between real space and momentum space.} We show the bipartite entanglement entropy as a function of time, measured for a cut next to the center of the wavepacket. These simulations were done at $D=320$ with a real-space window MPS of size $N_w=257$.}
\label{fig:wp}
\end{figure}
\par\noindent\emph{\textbf{Interpolating between spaces---}} %
We can explicitly show that momentum states develop less entanglement than real-space states, by considering wavepacket states that interpolate between these two limits. We prepare a Gaussian wavepacket centered around momentum $q$
\begin{equation}
\ket{\Psi_{q,\eta}^\alpha} = \sum_n \int \frac{dq'}{2\pi} \e^{-\frac{\eta}{2}(q-q')^2+iq'n} S_n^\alpha \ket{\Psi_0},
\end{equation}
and evolve it through time using a real-space window MPS as described above. In Fig.~\ref{fig:wp} we plot the evolution of the real-space bipartite entanglement entropy at the center of the wavepacket, showing that (i) the initial entanglement of a real-space localized wavepacket is smaller, but (ii) the increase of entanglement through time is a lot larger than for a wavepacket that is strongly localized in momentum space. The larger initial entanglement occurs because we are representing a momentum superposition in real space, but this zero-time offset can be transformed away by performing all calculations in momentum space directly. This result, therefore, confirms our previous observation that a momentum-resolved time evolution requires a smaller bond dimension for the same accuracy.

\par
\begin{figure}
\includegraphics[width=\columnwidth]{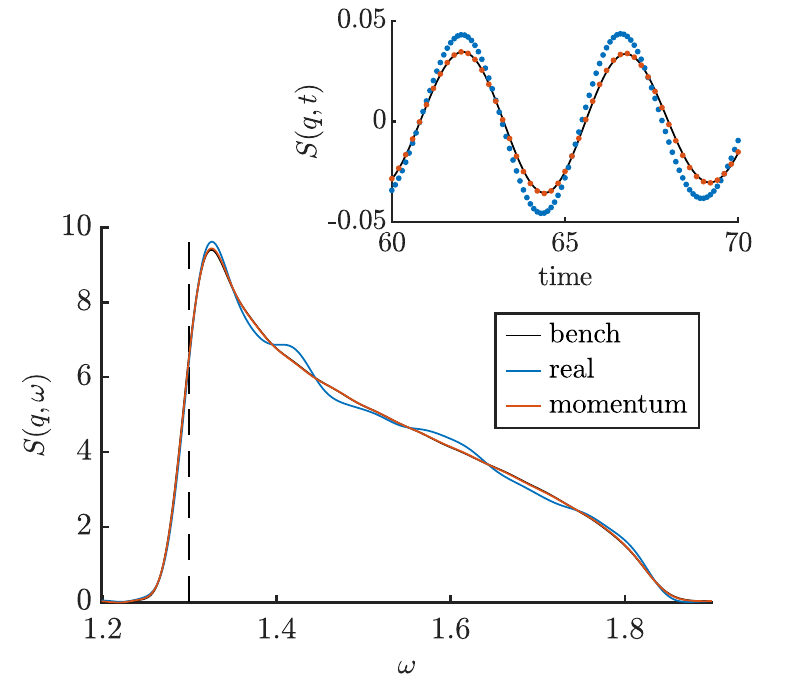}
\caption{\textbf{Spectral function for the XXZ chain.} We compare real- and momentum-space computations for $S(q,\omega)$ at the same bond dimension $D\approx100$, with a window size of $N_w=257$ and $N_w=64$ for the real- and momentum-space windows, respectively. We have evolved for a total time of $T=128$ and used a Gaussian envelope with $\alpha=3$. The vertical line is the edge of the continuum known exactly from the Bethe ansatz. The inset shows a portion of the real-time signals. We compare with a benchmark real-space simulation with bond dimension $D=1568$. We have used U(1) symmetry and a two-site uniform MPS in these simulations.}
\label{fig:xxz}
\end{figure}

\noindent\emph{\textbf{Benchmarks---}} %
Let us now perform some benchmarks on more challenging models. First we take the spin-1/2 XXZ chain, defined by the Hamiltonian
\begin{equation}
    H_{\mathrm{XXZ}} = \sum_{\braket{ij}} S^x_iS^x_j + S^y_iS^y_j + \Delta S^z_iS^z_j,
\end{equation}
where $S^\alpha_i$ are now spin-1/2 operators. We take $\Delta=1/2$ in the gapless phase, and investigate a cut of the spectral function at momentum $q=\pi/2$. It is known from Luttinger-liquid theory \cite{Imambekov2012} that the spectral function diverges at the edge of the continuum, which has been confirmed by MPS methods and Bethe-ansatz techniques \cite{Pereira2008, Caux2012}. Moreover, the continuum consists of multi-spinon states \cite{Caux2012} and is, therefore, potentially more challenging to resolve using our approach. In Fig.~\ref{fig:xxz} we show a comparison between the real-space and momentum-space approaches, again showing that the latter is more precise for the same bond dimension. In particular, the inset shows that the momentum-space window can capture the exact time evolution for significantly longer times.

\begin{figure}
\includegraphics[width=\columnwidth]{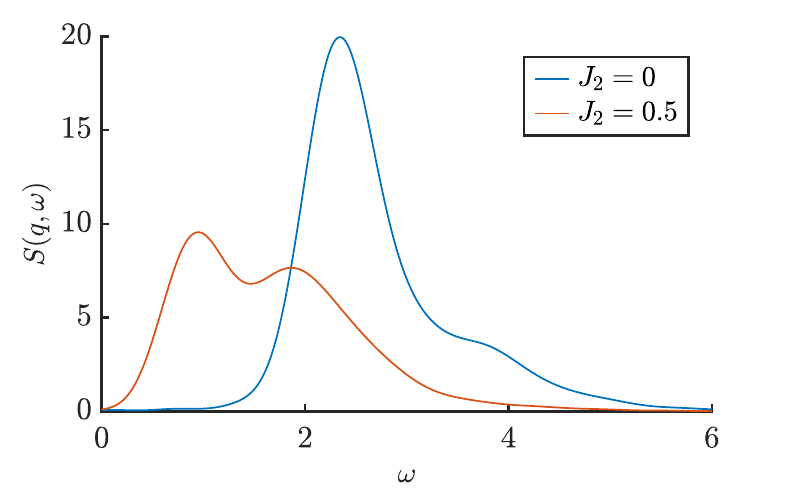}
\caption{\textbf{Spectral function for the $J_1$-$J_2$ model for momentum $q=(\pi,0)$ on a six-leg cylinder.} We have evolved until total time $T=9.5$ and applied a Gaussian envelope with $\alpha=4$; in both cases we have used SU(2) symmetry, a bond dimension around $D\approx400$ and a window size of $N_w=24$.}
\label{fig:j1j2}
\end{figure}

\par Finally, we consider the $J_1$-$J_2$ Heisenberg model on the square lattice, 
\begin{equation}
H = J_1 \sum_{\braket{ij}} \vec{S}_i \cdot \vec{S}_j + J_2 \sum_{\braket{\braket{ij}}} \vec{S}_i \cdot \vec{S}_j.
\end{equation}
Determining the phase diagram of the full two-dimensional model has been the subject of many numerical works, and getting accurate and unbiased results for the spectral function is important for getting an insight into the fractionalized spinon excitations in the putative spin-liquid phase. Even for the nearest-neighbor case ($J_2=0$) there is the occurrence of a dip in the magnon dispersion relation at momentum $(\pi,0)$ \cite{DallaPiazza2015} for which many physical mechanisms have been proposed \cite{Shao2017, Verresen2018, Ferrari2018b, Powalski2018}. This dip is expected to further drop in energy, until it becomes a gapless Dirac cone in the spin-liquid phase \cite{Hu2013, Ferrari2018b, Ferrari2021}.
\par Here we simulate the model on a six-leg cylinder and use our momentum-space approach to compute a cut of the spectral function for momentum $q=(\pi,0)$ for the unfrustrated case ($J_2=0$) and for $J_2/J_1=0.5$; in Fig.~\ref{fig:j1j2} the results are presented. For $J_2=0$ we find a gap that agrees with other methods \cite{Shao2017, Verresen2018, Ferrari2018b, Powalski2018, Vanderstraeten2019}. We clearly see that the gap has decreased significantly in the frustrated case, and that the spectrum is less peaked and more diffuse, which is suggestive of a fractionalized multi-spinon continuum. Nonetheless, we find that the gap is still quite large and we observe a two-peak structure in the lineshape, in quantitative agreement with a variational Monte-Carlo result in Ref.~\onlinecite{Ferrari2021}.  We have also computed the gap using the quasiparticle ansatz for cylinders \cite{VanDamme2021} and obtained a similar value (not shown), suggesting that this is a genuine feature of the six-leg cylinder.

\par\noindent\emph{\textbf{Conclusions---}} %
In this work we have introduced an MPS-based method for calculating spectral functions, relying on an efficient scheme for time-evolving momentum states. We have benchmarked our method against the traditional real-space approach and obtained more accurate results at relatively small bond dimensions. We expect that our method will be very useful in combination with a real-space approach: the latter allows us to find the full spectral function in a single run, whereas the former can be used for resolving fine-grained structures in certain momentum cuts. In particular, this strategy can be useful for tackling quasi two-dimensional systems with strong correlations such as spin liquids and Hubbard models, for which MPS simulations are currently limited to cylinders with small circumference or short evolution times \cite{Gohlke2017, Verresen2018, Verresen2019, Yang2019, Bohrdt2020, Kadow2022}.
\par In our work we have solely relied on time evolution for computing spectral functions, but we can also explore the momentum-space version of other strategies such as the correction-vector method. Note that previous works have indeed used (quasi-) momentum states on finite systems \cite{Kuhner1999, Dargel2011, Holzner2011, Xie2018, Wang2019, Kadow2022}, but without exploiting the structure of the momentum superposition directly in the thermodynamic limit. It would also be interesting to compare our work to a recent momentum-resolved method where time-dependent scattering events are simulated by time-dependent MPS methods \cite{Zawadzki2020}.
\par The ability to obtain high precision results at lower bond dimensions is of primary importance when applying projected entangled-pair states (PEPS), the two-dimensional version of MPS, where computational complexity scales unfavourably in bond dimension. Following the success of existing momentum-space methods for PEPS \cite{Vanderstraeten2015b, Vanderstraeten2019, Ponsioen2020}, we should be able to generalize our momentum-resolved scheme to higher spatial dimensions, and enable the calculation of high-precision spectral functions with PEPS.

\par\noindent\emph{\textbf{Acknowledgments---}} %
We acknowledge inspiring discussion with Markus Drescher, Jutho Haegeman, Wilhelm Kadow, Michael Knap, Ian McCulloch, Frank Pollmann, Frank Verstraete, and Elisabeth Wybo. This work was supported by the Research Foundation Flanders.

\bibliography{bibliography}

\appendix

\section*{Appendix}

\renewcommand{\diagram}[1]{\;\vcenter{\hbox{\includegraphics[scale=0.32,page=#1]{./diagrams_appendix.pdf}}}\;}

The class of momentum-window states on top of uniform MPS was introduced in Ref.~\onlinecite{Vanderstraeten2020}, where it was used to simulate extended bound states of solitons in a spin-1 model. In this Appendix, we will explain the properties of this class of variational states in some detail, and indicate how to simulate time evolution.

\par\noindent\emph{\textbf{Properties---}} %
We start from a uniform MPS (see Ref.~\onlinecite{Vanderstraeten2019b} for details), which can be represented in center-site gauge as
\begin{equation}
    \ket{\Psi_0} = \diagram{1},
\end{equation}
where $A_l$ and $A_r$ are left and right isometries. On top of this uniform MPS, a momentum-window state can be understood as the following state
\begin{equation}
 \sum_n e^{i q n} \diagram{2},
\end{equation}
where $n$ denotes the $n$'th site in the chain. This class of states is a linear subspace in terms of the tensor $B$, and has residual gauge freedom since the following subclass of states
\begin{multline}
 \sum_n e^{i q n} \left( \diagram{3} \right. \\  \left. - e^{i q} \diagram{4} \right)
\end{multline}
yields a zero norm. If we assume that the window state is orthogonal to the ground-state MPS, then we can absorb this gauge freedom entirely by constraining $B$ to be of the form
\begin{equation}
\diagram{5} = \diagram{6},
\end{equation}
with $V_l$ being the null space of $A_l$, satisfying
\begin{equation}
\diagram{7} = 0 , \qquad \diagram{8} = \diagram{9}.
\end{equation}
Working with such an extended block tensor the variational parameters scale exponentially with the size of the block, so we make the approximation of decomposing $X$ as a finite MPS
\begin{equation}
\diagram{5} = \diagram{10} ,
\end{equation}
so that we arrive at our variational class of momentum-window MPS or ``$q$-MPS'' as
\begin{equation} \label{eq:app_window}
    \ket{\Phi_q(X)} = \sum_{n}\e^{iqn} \diagram{11}.
\end{equation}
This class of $q$-MPS can be utilized in much the same way as a ``usual'' finite MPS. After all, the overlap of two different $q$-MPSs simplifies to the overlap between their respective windows:
\begin{multline} \label{eq:normX}
\braket{\Phi_q(X) | \Phi_{q'}(X')} \\ = 2\pi\delta(q-q') \braket{\Phi_q(X) | \Phi_{q}(X')}_\mathrm{finite}
\end{multline}
with
\begin{equation}
\braket{\Phi_q(X) | \Phi_{q}(X')}_\mathrm{finite} = \diagram{12} \cdots \diagram{13}.
\end{equation}
This means that the orthogonality center can be shifted through the window by consecutive orthogonal matrix decompositions, and we can therefore directly translate most algorithms from their real-space to the momentum-space setting!

\par\noindent\emph{\textbf{Energy minimization---}} %
As a first example, we can consider a direct energy minimization by sweeping through the window, similarly as the standard DMRG algorithm for ground states of a finite spin chain \cite{Schollwoeck2011}. Indeed, we can iteratively solve eigenvalue problems, pushing the orthogonality center at site $n$ around:
\begin{equation} \label{eq:Heff}
    (\Heff)_{nn} X_n = \lambda X_n.
\end{equation}
Because our ansatz has a well defined momentum and is by construction orthogonal to the ground state, we will be directly targeting the lowest-lying excitations. This algorithm was used in Ref.~\onlinecite{Vanderstraeten2020} for simulating dispersion relations of broad bound states in a spin chain.
\par The calculation of $\Heff$ is quite straightforward, it is defined as
\begin{equation}
    (\Heff)_{ij}  = \frac{\partial^2}{\partial \bar{X}_i \partial X_j} \bra{\Phi_q(X)} H \ket{\Phi_q(X)}_{\mathrm{finite}},
\end{equation}
where we are typically only interested in the diagonal entries: when sweeping through the window we only vary one tensor, see Eq.\eqref{eq:Heff}.
\par Here, the energy expectation value it is the momentum sum of all possible different positions of a window in the top and bottom layer. For two normalized $q$-MPS parametrized by $X$ and $X'$ and the Hamiltonian represented as a matrix product operator (MPO) \cite{Schollwoeck2011}, we would require the following sums:
\begin{align}
 & \bra{\Phi_q(X)} H \ket{\Phi_q(X)}_{\mathrm{finite}} \nonumber \\
 & = \sum_{n'>n} e^{i (n-n') q} \diagram{21} \cdots \diagram{22} \nonumber \\
 & + \sum_{n'<n} e^{i (n-n') q} \diagram{23} \cdots \diagram{24} \nonumber \\
 & + e^{i q} \diagram{25} \nonumber \\
 & + e^{i 2 q} \diagram{46} \nonumber \\
 & + e^{-i q} \diagram{26} \nonumber \\
 & + e^{-i 2 q} \diagram{45} \nonumber \\
 & + \diagram{27} .
\end{align}
Here, the first two sums are over $n$, whereas the index $n'$ is an arbitrary site in the lattice; the prescription $\braket{\cdots}_{\mathrm{finite}}$ denotes that we have taken out one infinite sum that occurs when taking overlaps of momentum states, see Eq.~\eqref{eq:normX}. The first two sums contain an infinite amount of terms, and can be summed by solving the linear problem:
\begin{equation}
\diagram{28} \left( 1 - \e^{-iq} \diagram{29} \right)^{-1}.
\end{equation}
This type of infinite summation always occurs when working with momentum states on top of uniform MPS, and details can be found in Ref.~\onlinecite{Vanderstraeten2019b}. This linear problem has to be recomputed every time that any $X_n$ changes, which makes some sweeping algorithms more expensive compared to their finite MPS counterparts. Importantly, however, if a tensor in the bra changes we do not have to recompute these contributions!

\par\noindent\emph{\textbf{Initial state---}} %
In the main text we have looked at real-time evolution within the class of momentum-window MPS, where the initial state is a momentum operator on the ground state. The initial state is of the form
\begin{equation}
    \sum_n \e^{iqn} O_n \ket{\Psi_0} = \diagram{14},
\end{equation}
which can be brought under the form of Eq.~\ref{eq:app_window} by maximizing the overlap with a two-site window. Since this optimization problem is linear in $X_1$, the solution is simply
\begin{equation}
    X_1 = \frac{\partial}{\partial \bar{X_1}} \bra{\Phi_q(X_1)} \sum_n \e^{iqn} O_n \ket{\Psi_0} 
\end{equation}
The overlap in question is given by the terms
\begin{multline}
    \diagram{15}  + \e^{-iq} \diagram{16} \\ + \e^{-2iq} \diagram{17} + \cdots,
\end{multline}
where the latter terms can be summed as
\begin{equation}
    \e^{-iq} \diagram{18} \left( 1 - \e^{-iq} \diagram{19} \right)^{-1} \diagram{20}.
\end{equation}
Again, this type of infinite summation is usual when working with momentum states on top of uniform MPS \cite{Vanderstraeten2019b}. 
\par After having reformulated the initial state as a $q$-MPS, we can perform the actual time evolution. Here, we can choose between applying the time-dependent variational principle (TDVP) \cite{Haegeman2011, Haegeman2016} or applying an MPO representation for the time-evolution operator with variational truncation.

\par\noindent\emph{\textbf{TDVP---}} %
For applying the TDVP, we require the same $\Heff$ as in $q$-MPS enerhy minimization. The resulting algorithm is identical to the well known finite-MPS variant as in Ref.~\onlinecite{Haegeman2016}, where for every time step $\delta t$ one sweeps through the tensors and evolves the tensors as: (i) evolve the current orthogonality center as
\begin{equation}
    \e^{ - \frac{i \delta t}{2} \Heff }(X_i)  = X_i',
\end{equation}
(ii) using QR or RQ decompositions we can then left- or right-orthogonalize $X_i'$, (iii) subsequently evolve the resulting $R$ backwards in time,
\begin{equation}
    \e^{ - \frac{i \delta t}{2} \Heff}(R)  = R',
\end{equation}
and (iv) absorb $R'$ in the next tensor. Note that the exponentials can be implemented iteratively such that only the action of $\Heff$ is needed on a given tensor; the action of $\Heff$ on the tensor $X_i$ is the same as the one in Eq.~\eqref{eq:Heff}, whereas the action on a link matrix $R$ can be found by projecting the same $\Heff$ \cite{Haegeman2016}. After a full forward and backwards sweep, we have evolved the entire window forward with one time step $\delta t$. 
\par We note that this algorithm is rather expensive, as every exponentiation step has to be done iteratively and $\Heff(X)$ can re-use very little information between applications on the different $X_i$ in the window. This is in contrast to the usual finite-MPS setting where $\Heff(X)$ can be updated by a single contraction to the next site in the chain.

\par\noindent\emph{\textbf{MPO time evolution---}} %
Applying an MPO to a $q$-MPS goes as follows. In each time step, we want to update the $X$ tensors as
\begin{equation}
    \ket{\Phi_q(X')} \approx U(\delta t) \ket{\Phi_q(X)},
\end{equation}
where the $\approx$ sign can be interpreted variationally as minimizing the norm between these two states. Solving this minimization problem can be done, again, by a sweeping algorithm, where we sweep over the $X'$ tensors until convergence. By shifting the orthogonality center accordingly, the minimization problem for a single $X'$ tensor is simply solved as
\begin{equation} \label{eq:update_Xi}
    X_i' = \frac{\partial}{\partial \bar{X'_i}} \bra{\Phi_q(X')}  U(\delta t) \ket{\Phi_q(X)}_{\mathrm{finite}}.
\end{equation}
This quantity is a sum of different terms: Because the MPO is between bra and ket layer, one does not have the nice property that only one term survives as in Eq.~\eqref{eq:normX}. The term where the two windows are located on the same region is proportional to
\begin{equation} \label{eq:onsite}
    \diagram{30},
\end{equation}
where the $U$ tensor encodes the MPO representation of the time-evolution operator, and with $G_l$ the fixed point of the channel
\begin{equation}
    \diagram{31} = \lambda \diagram{32},
\end{equation}
and a similar equation for $G_r$. If the uniform MPS is close to the ground state and $U(\delta)$ is close to an exact representation of $\e^{-i(H-E_0)\delta t}$, then the eigenvalue $\lambda$ is close to one. It is convenient to explicitly rescale the MPO tensor for $U(\delta t)$ such that this eigenvalue is exactly equal to one. We also require the fixed points to be normalized as
\begin{equation}
    \diagram{33} = 1.
\end{equation}
Then the above diagram in Eq.~\eqref{eq:onsite} also appears as a term without a prefactor.
\par Other terms are of the form
\begin{equation}
    \diagram{34},
\end{equation}
and should all be summed. The infinite number of disconnected terms can be explicitly summed by solving a linear problem:
\begin{multline}
    \diagram{35} \left( 1- \e^{-iq} \diagram{36} \right)^{-1} \\ \diagram{37},
\end{multline}
as is usual for working with momentum states on uniform MPS, see Ref.~\onlinecite{Vanderstraeten2019b} for details. Summing up all these terms allows us to compute the expression in Eq.~\eqref{eq:update_Xi} and to update the tensor $X_i'$.
\par Note that in the course of sweeping through the window, the different contributions to Eq.~\eqref{eq:update_Xi} do not need to be computed from scratch, but the environments can be updated by a single contraction. This is very similar to a real-space window optimization, only are there a number of different terms to take into account. These updates can be performed in parallel, so with enough parallel workers the optimization in each time step does not need to take more time than the real-space approach. Given that we can reduce the bond dimension significantly in momentum space (as shown in the main text), the momentum-space approach is more efficient for obtaining a single momentum cut of the spectral function.

\par\noindent\emph{\textbf{Larger unit cells---}} %
Until now we have explained everything for uniform MPS with a single-site unit cell, but larger unit cells often need to be considered, particularly if we consider multi-leg ladders or cylindrical geometries. The extension to larger unit cells is mainly a matter of more intricate bookkeeping, but it is useful to write down the expressions for a $q$-MPS. We will describe the case of two-site unit cells, larger unit cells follow trivially.
\par A two-site uniform MPS is given by
\begin{align}
    \ket{\Psi_0} &= \diagram{38} \\
    &=  \diagram{39}.
\end{align}
As explained in Ref.~\onlinecite{ZaunerStauber2018}, a momentum state on top of such an MPS is given by
\begin{multline}
    \sum_{n\in2\Z}\e^{iqn} \left( \diagram{40} \right. \\ + \left. \diagram{41} \right),
\end{multline}
i.e. we take two different $B$ tensors for each of two sites within the unit cell. The extension of this to a momentum-window state is of the form
\begin{align}
    & \ket{\Phi_q(X^1,X^2)} = \sum_{n\in2\Z}\e^{iqn} \nonumber \\ 
     & \left( \diagram{42} \right. \nonumber \\
    & + \left. \diagram{43} \right) .
\end{align}
Note that an alternative choice would consist of taking a single window that represents the sum of the two windows, but this would require a larger bond dimension. In particular, a momentum operator can be represented using the above expression without changing the bond dimension, whereas representing this with a single window would require us to double the bond dimension within the window.
\par The definition of the $q$-MPS with inclusion of the null-space projectors $V_l^i$ makes sure that the inner product remains of a simple euclidean form
\begin{multline}
    \braket{\Phi_{q'}(X'^1,X'^2)|\Phi_q(X^1,X^2)}_\mathrm{finite} \\ = \sum_i \diagram{44}.
\end{multline}
Therefore, a sweeping algorithm can still be used for energy minimization and MPO time evolution. For the former, the effective hamiltonian does not diagonalize in the different windows and cross terms need to be taken into account. For the MPO time-evolution, however, the sweeping procedure over the tensors in each window is completely independent from the others, so we can parallellize these completely!

\end{document}